\documentclass[final,5p,times,twocolumn]{elsarticle}

\usepackage{amssymb}

\journal{Astroparticle Physics}

\begin{document}

\begin{frontmatter}



\title{Dethinning Extensive Air Shower Simulations}


\author[label1]{B.T.~Stokes\corref{cor1}}
\ead{stokes@cosmic.utah.edu}

\author[label1]{R.~Cady}

\author[label1,label2]{D.~Ivanov}

\author[label1]{J.~N.~Matthews}

\author[label1]{G.~B.~Thomson}

\address[label1]{University of Utah, Department of Physics \& Astronomy and High Energy Astrophysics Institute, Salt Lake City, Utah 84112, USA}
\address[label2]{Rutgers---The State University of New Jersey, Department of Physics and Astronomy, Piscataway, New Jersey 08854, USA}
\cortext[cor1]{Corresponding author}

\begin{abstract}

We describe a method for restoring information lost during statistical
thinning in extensive air shower simulations.  By converting weighted particles
from thinned simulations to swarms of particles with similar
characteristics, we obtain a result that is essentially identical to
the thinned shower, and which is very similar to non-thinned
simulations of showers.  We call this method dethinning.  Using
non-thinned showers on a large scale is impossible because of
unrealistic CPU time requirements, but with thinned showers that have
been dethinned, it is possible to carry out large-scale simulation
studies of the detector response for ultra-high energy cosmic ray
surface arrays.  The dethinning method is described in detail and
comparisons are presented with parent thinned showers and with
non-thinned showers.

\end{abstract}


\end{frontmatter}


\section{Introduction}
\label{sect:intro}

In the study of ultrahigh energy cosmic rays (UHECR) two main
experimental techniques have been used: detection of the fluorescence
light emitted by nitrogen molecules excited by the passage of
particles in extensive air showers, and detection of the particles
themselves when they strike the ground by deploying an array of
particle detectors over a large area.  Analysis of data from a
fluorescence detector involves making a detailed Monte Carlo
simulation of the shower, atmosphere, and detector \cite{hires_fadc}
\cite{auger:endethybrid}.  Only by this technique can the aperture be
calculated as a function of energy.  Simulation of a large number of
complete showers can not be performed using programs like CORSIKA
\cite{corsika} or AIRES \cite{aires} because the CPU requirements are
too large.  The approximation technique called thinning is therefore
used, in which particles are removed from consideration in the shower
generation and other particles in similar regions of phase space are
given weights to account for the loss.  Since a fluorescence detector
is sensitive to the core region of a shower, where $10^{11}$ charged
particles occur at shower maximum for a $10^{20}$ eV event, thinning
does not harm the accuracy of the simulation.  However, for an array
of surface detectors (SD), where several km from the core the density
of particles is low, the thinning approximation produces an
unacceptably coarse simulation of a shower.  The average density of particles, 
as a function of radius from the core, in a thinned shower is approximately 
correct, but the fluctuations about the average are much larger than in a 
shower generated without using the thinning approximation.  The result of 
this is that the Monte Carlo technique has been available to those analyzing
SD data only in a very limited way.

We have developed a method of performing an accurate Monte Carlo
simulation of the surface detector of the Telescope Array (TA)
experiment.  This method consists of three parts: (1) generating ~100
non-thinned CORSIKA showers above $10^{18.5}$ eV using many computer
nodes operating in parallel \cite{parallel}, (2) using these showers
to develop a method (called dethinning) of replacing much of the shower
information lost in thinning, and (3) generating a large number of
dethinned showers, including a detailed simulation of the TA surface
detector performance, and comparing histograms of important quantities
between the data and the Monte Carlo simulation.  Reference \cite{parallel}
describes a method for generating CORSIKA showers using many computing
nodes in parallel.  The present work describes
the second step of the method: dethinning.  A future paper will
describe the third step.  This method has proved quite successful, and
has allowed us to calculate the aperture of the TA surface detector
even in the energy range where its trigger efficiency is not 100\%.

The idea of replacing the information lost in thinning was first
introduced by P. Billoir \cite{billoir:dth}.  The basic idea in reference 
\cite{billoir:dth} and in our work is the same:  start with a thinned shower, 
maintain the average density of particles, and smooth the distribution to 
get the correct amount of fluctuations.  He considered CORSIKA output 
particles striking the ground in the vicinity of a surface detector
(specifically a detector of the Pierre Auger experiment), and, by an
oversampling technique based on the CORSIKA output particles, predicted what 
that detector should observe.  He named this technique unthinning.  He 
continued on to estimate several systematic biases to which his method might 
be sensitive, by studying the properties of thinned and unthinned showers.  

Our method, as described in this paper, differs in two ways.  First, as the 
dethinning process we take each CORSIKA output particle with weight ${w}$ and 
from it generate a swarm of ${w}$ particles.  We perform this step for the 
entire set of CORSIKA output particles.  The details of this generation 
matter, and are described in this paper.  Second, we compare the resulting 
dethinned showers with showers of almost identical characteristics generated 
without using the thinning approximation.  This is a direct way of testing 
the accuracy of the dethinning process.  Since real surface detectors for 
ultrahigh energy cosmic rays sample showers coarsely, and measure only the 
time distribution of the number of particles that strike them, this is the 
important aspect of the comparison process.  We present our comparisons 
between dethinned showers and non-thinned showers in this manner.

\section{Dethinning Description}
\label{sect:parallel}

In a thinned EAS simulation, particles are discarded from the simulation in 
order to conserve computation time.  In the case of CORSIKA 
\cite{corsika}, 
for a given thinning level, 
$\varepsilon_{th}$, if the energy sum of all $j$ secondary 
particles falls below the thinning energy
\begin{equation}
\varepsilon_{th}E_0>\sum_jE_j.
\end{equation}
then only one randomly assigned secondary particle survives with probability
\begin{equation}
p_i=E_i/\sum_jE_j.
\end{equation}
If the energy sum is greater than the thinning energy, then secondary 
particles with energy below the thinning energy survive with probability
\begin{equation}
p_i=E_i/(\varepsilon_{th}E_0).
\end{equation}
In both cases, surviving particles have their weight multiplied by a factor of
$w_i=1/p_i$.  Thus the weight of a particle reaching the end of the simulation
after passing through $k$ vertices is
\begin{equation}
w_{i,tot}=\prod_k1/p_k.
\end{equation}

For sufficiently low values of $\varepsilon_{th}$, it is clear that the thinned
simulation output can be thought of as an accurate sample of secondary particle 
types, trajectories, and positions compared to a non-thinned 
simulation, for the observable parts of the shower.  In this situation, for a 
particle of weight, $w_i$, 
the simulation, on average, removed $w_i-1$ particles from a similar position 
in phase space. 
(Of course, if the value of $\varepsilon_{th}$ is increased, this situation
is no longer valid because the thinned simulation no longer has the same 
distribution of particle types, trajectories, and energies
as the parent shower.)  By comparing a dethinned shower with a similar 
non-thinned shower one can determine the accuracy of the sampling.   The 
pivotal questions are then: 
(1) How can a thinned sample be used to reconstruct the full simulation?\ and 
(2) What is the maximum value of $\varepsilon_{th}$ for which the thinned 
sample accurately represents the parent shower's particle types, etc.?

\begin{figure}[t,b]
  \centering
  \includegraphics[width=0.48\textwidth]{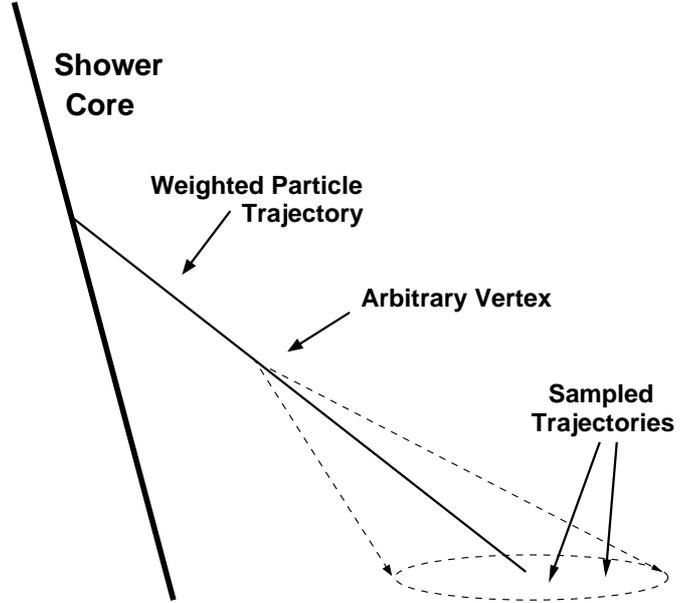}
  \caption{Geometry for a ``Gaussian cone'' with a vertex placed at arbitrary
position on the trajectory of the weighted particle}
  \label{fig:1}
 \end{figure}

We address the first question by describing the process by which we dethin the 
showers.  The original CORSIKA shower consists of a list of output particles 
(plus their weights, types, energies, positions, angles, and arrival time) 
that have struck the ground.  Dethinning consists of adding particles to this 
list.  For every CORSIKA output particle of weight $w$ we add $w-1$ particles 
to the list.  When this is completed the weight of each particle is set to $1$. 
To insert these particles we use the following procedure (see Figure~\ref{fig:1}).
\begin{enumerate}
\item Choose a vertex point on the trajectory of the weighted particle, in the 
way given in the next paragraph.
\item Choose a point in a cone centered on the output particle's trajectory, 
weighted by a two-dimensional Gaussian distribution with a sigma of a few 
degrees (as described in Section 3).  This will be the inserted particle's 
trajectory.
\item Project the inserted particle to ground level, assign it a time and 
energy (as described in Section 3), and add it to the particle list of the 
dethinned shower.
\item Perform steps 2-3 $w-1$ times.  For the case where $w$ is not an
integer, add one particle randomly based on the decimal part of $w$.
\end{enumerate}

\begin{figure}[t,b]
  \centering
  \includegraphics[width=0.48\textwidth]{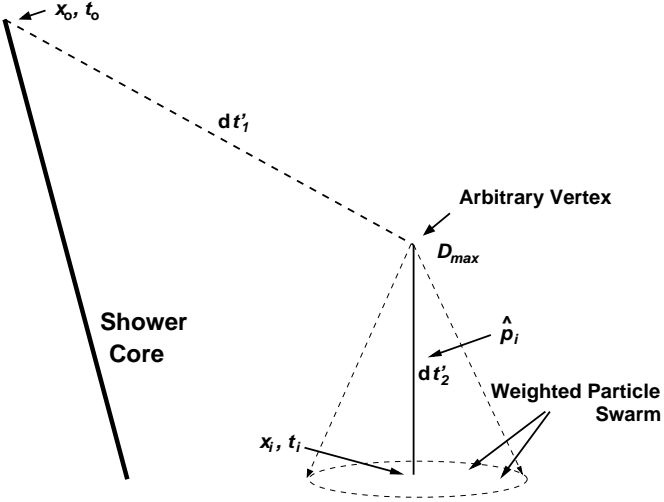}
  \caption{In order to ensure temporal consistency in the EAS simulation, 
    we require
    $t_i-t_0 \geq {\rm d} t_1'+{\rm d}t_2'$, where $t_i$ is 
    the recorded arrival time for weighted particle and $t_0$ is the time of 
    first interaction.}
  \label{fig:2}
 \end{figure}

There is a maximum distance from the ground that one can choose for the vertex 
in item 1 above, which is set by the requirement that no particle can have an 
arrival time that precedes the arrival of the shower front.  A too-early 
arrival time occurs when the total time-of-flight from the point of 
first interaction, ${\bf x_0}$, to the vertex point and 
then to the position on the ground of the 
generated particle is less than the time-of-flight directly from ${\bf x_0}$ to
final particle position.  This can be corrected by fixing
the position of the vertex point to a position where the time-of-flight 
from ${\bf x_0}$ to the imaginary vertex and then onward to the final position 
of the weighted particle, ${\bf x_i}$ is equal to the difference in the arrival
time of the weighted particle, $t_i$, and the time of first interaction, $t_0$.
This condition is:  the distance 
along the weighted particle trajectory, ${\bf\hat{p_i}}$, between ${\bf x_i}$
and the vertex point is
\begin{equation}
D_{max} = \frac{c^2(t_i-t_0)^2-|{\bf x_i-x_0}|^2}{2(c(t_i-t_0)-{\bf (x_i-x_0)}\cdot{\bf \hat{p_i}})},
\end{equation}
where $c$ is the speed of light.  We should emphasize that $D_{max}$ is the 
maximum separation between the vertex point and the ground.  Any 
shorter separation will also generate dethinned particles that are temporally 
consistent.
The calculation of $D_{max}$ is graphically illustrated in Figure~\ref{fig:2}.

The second important question pertains to selecting a value of
$\varepsilon_{th}$. It should be recalled that in order for this algorithm to
work, $\varepsilon_{th}$ must be sufficiently small so that the thinned 
simulation contains a large enough sample of output particles that the 
distribution of particle types, trajectories, and positions is the same as in 
the non-thinned simulation.  We addressed this question phenomenologically by 
comparing non-thinned and dethinned showers.  
Our conclusion was that for $\varepsilon_{th}=10^{-7}$, this algorithm could be
applied without further adjustment.  Conversely, for $\varepsilon_{th}=10^{-5}$,
the thinned sample did not contain enough output particles to successfully 
mimic the properties of a non-thinned shower.  The option in between, 
$\varepsilon_{th}=10^{-6}$, proved to be a borderline case.  We 
found that dethinning could be successfully implemented by careful adjustment 
of the available
parameters.  Because $10^{-6}$ showers take 1/10 the time to generate as 
$10^{-7}$ showers, and require 1/10 as much storage space, 
$\varepsilon_{th}=10^{-6}$ 
proved to be highly desirable so we chose to focus our efforts there. 

\section{Adjusting the Dethinning Algorithm}
\label{sect:tuning}

In adjusting the dethinning algorithm, we have sought agreement between
dethinned and non-thinned simulations for all measures relevant to observation
by the TA surface array. These measures include: distributions of secondary 
particle position and time, particle type,  incident angle, and energy.  
These measures were tested by comparing thinned versus dethinned
versions of the same simulation and then subsequently comparing dethinned 
simulations versus non-thinned simulations with identical input parameters.
Tuning was not considered complete until all measures agreed for lateral 
distances $[500,4500]$~m from the shower core.

In the first step, secondary particle spectra for thinned and dethinned versions
of the same shower were compared with respect to particle type (photons, 
electrons, and muons), incident angle with respect to the ground, and position
within the shower footprint.  The algorithm is tuned so that the particle 
fluxes after dethinning were consistent with those seen in the original thinned 
output.  

In the second step, distributions of particle fluxes over $6 \times 6$~m$^2$
detector-size areas are compared for dethinned and non-thinned
simulations.  For this purpose, a library of more than 100 non-thinned showers
was generated with parallelized CORSIKA~\cite{parallel}.  
This library contains showers
with $E_0=[10^{18.5},10^{19.5}]$eV, $\theta_0=[0,60]^\circ$, and proton and iron
primary particle types.
When identical input parameters are used for thinned and 
non-thinned simulations, the resulting simulations are not identical.  It is 
therefore necessary to normalize the net secondary particle fluxes of the 
non-thinned simulation with respect to the thinned simulation.
Once this normalization is accomplished the dethinning algorithm can be 
further refined so that shower particle fluctuations are consistent between 
dethinned and non-thinned simulations.  A further check is done by 
simultaneously examining dethinned versus non-thinned comparisons over 
many simulations without normalizing the non-thinned showers.  By utilizing
the 100 non-thinned showers for this comparison, 
we ensure that the thinning-dethinning process does not
bias the energy scale with respect to non-thinned showers.

The result of adjusting the parameters of the dethinning process is as follows:
\begin{enumerate}
\item Angle subtended by Gaussian cone:  Set to $\beta d$ where $d$ is the 
lateral distance from the shower core for the weighted particle and 
$\beta=3^\circ$/km for electromagnetic particles and $1^\circ$/km 
for muons and hadrons. 
The values of $\beta$ are the minimum 
necessary to dethin simulations with $\varepsilon_{th}=10^{-6}$.  A smaller
value of $\varepsilon_{th}$ enables the use of smaller $\beta$ values.
\item Energy perturbation:  Vary the energy of each particle in swarm about
a $\pm 10\%$ fractional Gaussian distribution centered on the energy of the 
original particle.  This correction smooths the secondary particle spectra.
\item Minimum lateral distance:  Since any detector within a few hundred 
meters of the shower core is saturated by the large number of particles in 
that part of a shower, it is not necessary to simulate the central part of a 
shower; not doing so saves a great deal of CPU time as well.  We therefore do 
not consider in the dethinning process any CORSIKA output particle within a 
distance $r_{min}$ of the core, and to avoid biases retain only particles 
farther than $r_{min}'$ from the core.  For the case of 
$\varepsilon_{th}=10^{-6}$, $r_{min}\geq 100$~m, and $r_{min}'-r_{min}\geq 200$~m.
\item Particle acceptance:  Some particles in the swarm with longer trajectories 
than the original weighted particle, if they were followed in the simulation, would not reach the ground.  We therefore introduce an acceptance for particles in the swarm with probability: 
$P=e^{-\Delta\chi/\epsilon}$, where $\Delta\chi$ is the difference in slant 
depth between the trajectories and $\epsilon=50$~g/cm$^2$.  This is an appreciable correction only for showers with large zenith angle.
\item Height of Gaussian cone:  When one sets the vertex a height $D_{max}$ above the ground, this aligns the time at the ground for reintroduced particles, but muons that are created late in the shower then have too wide a spatial distribution  A solution is to set the vertex distance to the smaller of $D_{max}$ and
\begin{equation}
D'=|{\bf x_i}-{\bf x_0}|-X^{-1}({\bf x_i},{\bf x_0},\alpha h),
\end{equation}
where ${\bf x_0}$ is the point of first interaction, $h$ is the generation
of the hadron 
 from which the particle originated, $\alpha = 30$~g/cm$^2$, and
$X^{-1}({\bf x_i},{\bf x_0},\alpha h)$ is the distance equivalent of 
$\alpha h$ slant depth on the trajectory from ${\bf x_0}$ to 
${\bf x_i}$.  
\end{enumerate}

\section{Comparing Simulations: An Example}
\label{sect:corsika}

In Section~\ref{sect:tuning} the method used to tune
the dethinning algorithm for $\varepsilon_{th}=10^{-6}$ thinned showers was
described.  We now consider two examples of comparisons between the dethinned result and the parent thinned shower and with a very similar non-thinned shower.
For the comparisons, we use CORSIKA v6.960~\cite{corsika}.  High energy hadronic interactions are modeled by QGSJET-II-03~\cite{qgsjet}, low energy hadronic interactions are modeled by FLUKA2008.3c~\cite{fluka1}\cite{fluka2}, and 
electromagnetic interactions are modeled by EGS4~\cite{egs4}.  For the thinned
shower, $\varepsilon=10^{-6}$  and additionally, we apply the thinning 
optimization scheme proposed by Kobal~\cite{kobal}. 

For both comparisons, the shower footprint is divided into eight 
500~m thick ring-shaped segments from 500 to 4500~m in lateral 
distance.  Each lateral ring is further divided into six pie-shaped wedges with 
respect to the rotation angle about the shower axis.
  
\begin{figure*}[t,b,p]
\begin{center}
(a)\includegraphics[width=0.975\textwidth]{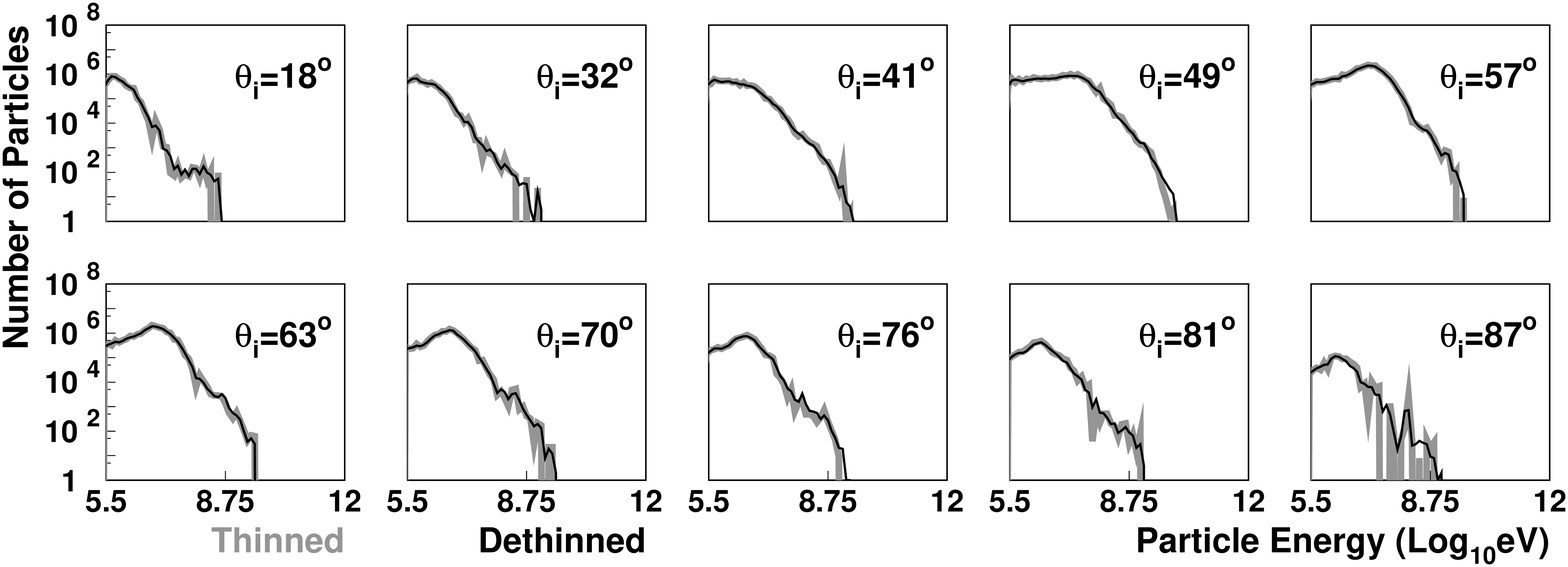}\\
(b)\includegraphics[width=0.975\textwidth]{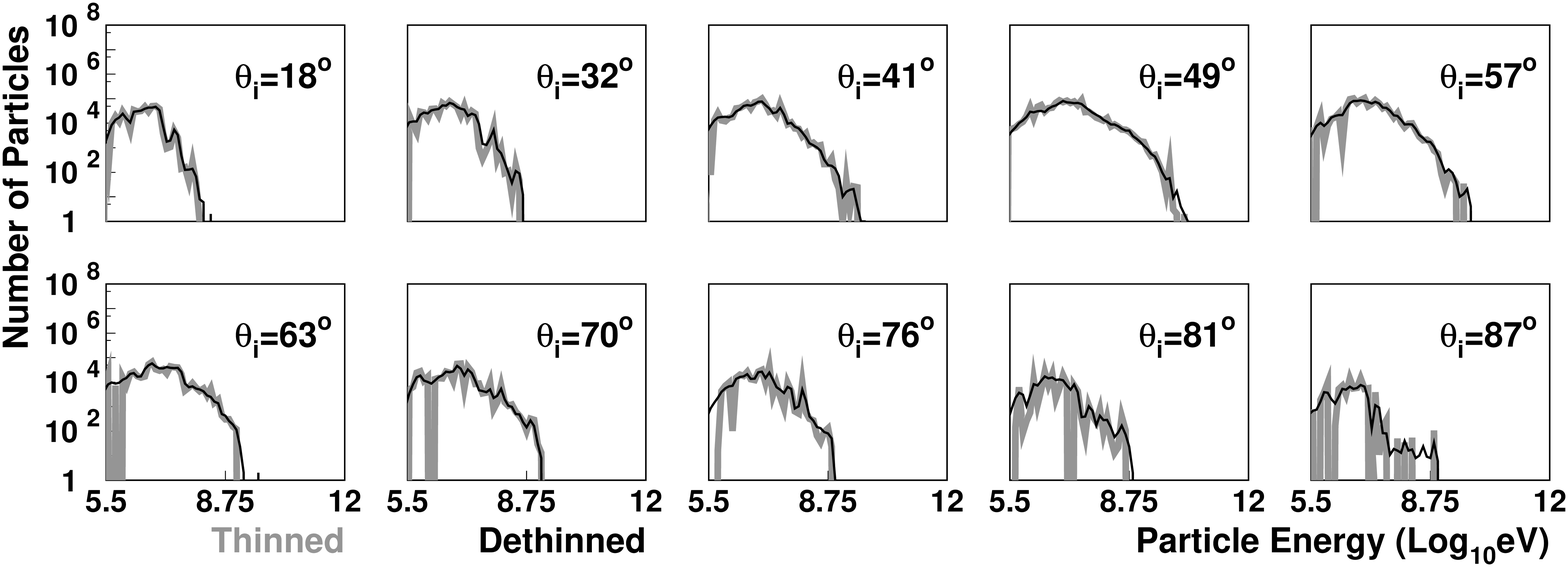}\\
(c)\includegraphics[width=0.975\textwidth]{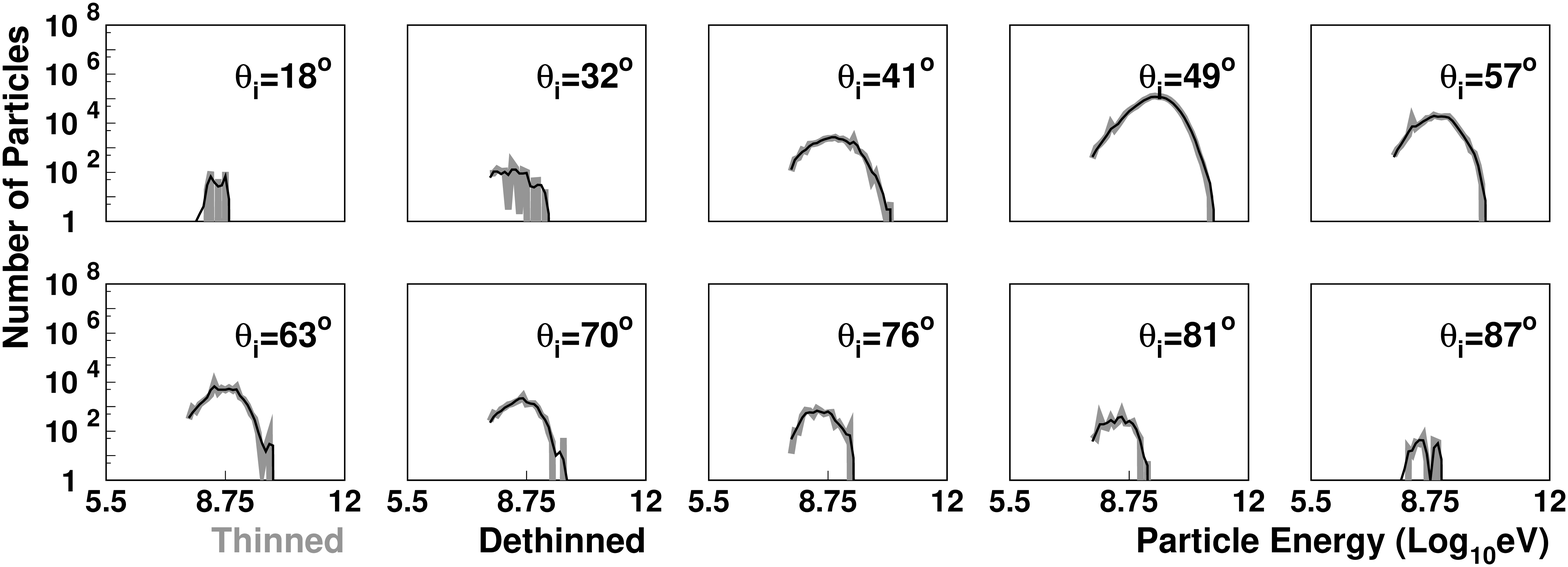}
\end{center}
\caption{Comparison of secondary particle spectra with and without dethinning 
for a thinned simulation of a protonic EAS with primary energy $E_0=10^{19.0}$~{eV} and primary zenith angle $\theta_0=45^\circ$.  In both cases, 
the secondary particles whose ground position was within
a region enclosed by shower rotation angles, $\Phi=[-30^\circ,30^\circ]$ (with
respect to the primary azimuthal direction) and 
lateral distances, $r=[500{\rm m},1000{\rm m}]$ were tabulated according
to particle type, incident angle with respect to the ground, $\theta_i$, and 
kinetic energy.  In the case of the thinned simulation, each secondary particle 
with weight, $w_i$, was treated as $w_i$ identical particles. 
The resulting spectral comparisons are shown in $\cos\theta_i=0.1$ increment
bins for a) photons, b) electrons, and c) muons. For each 
histogram, good agreement is observed between thinned simulations (gray) and 
dethinned (black).}
\label{fig:thdth1}
\end{figure*}

\begin{figure*}[t,b,p]
\begin{center}
(a)\includegraphics[width=0.975\textwidth]{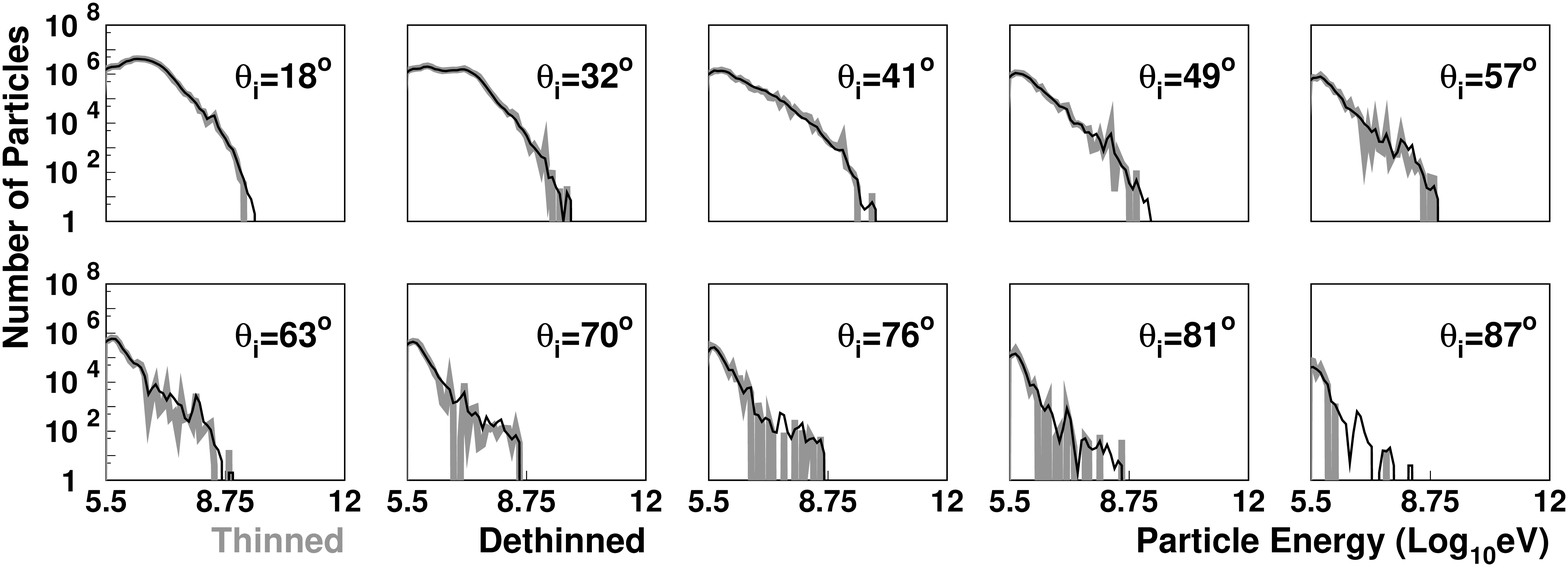}\\
(b)\includegraphics[width=0.975\textwidth]{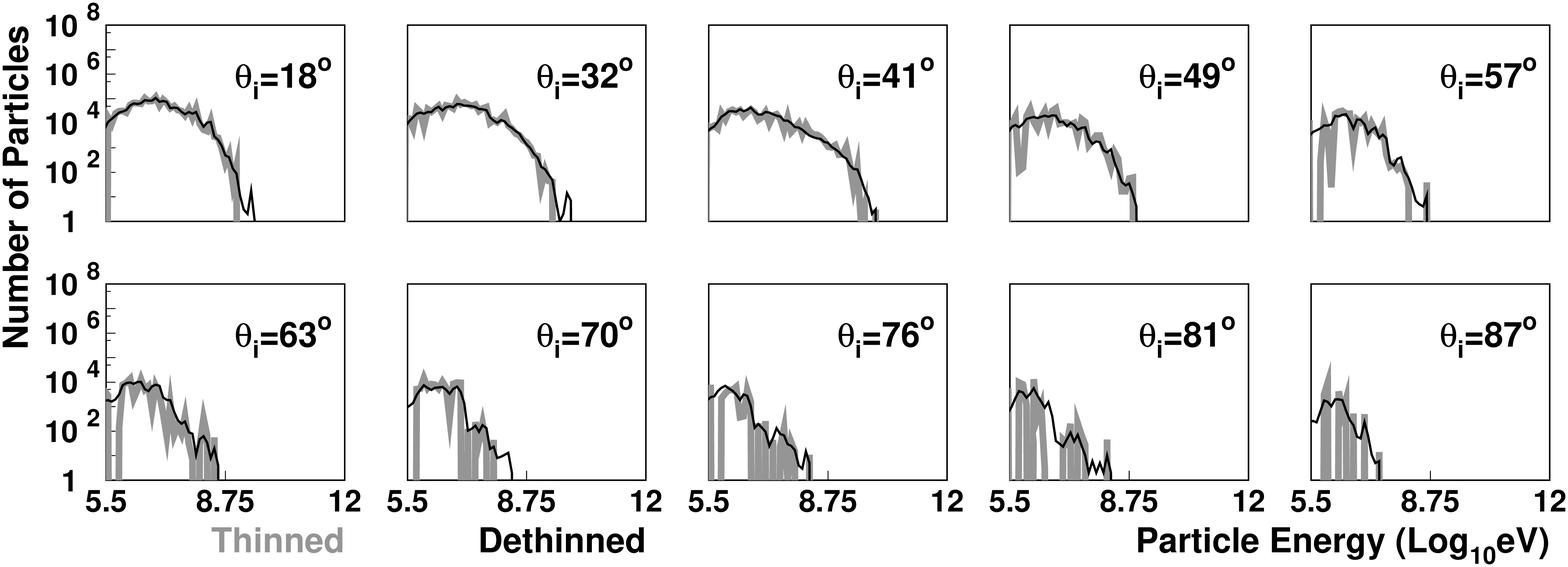}\\
(c)\includegraphics[width=0.975\textwidth]{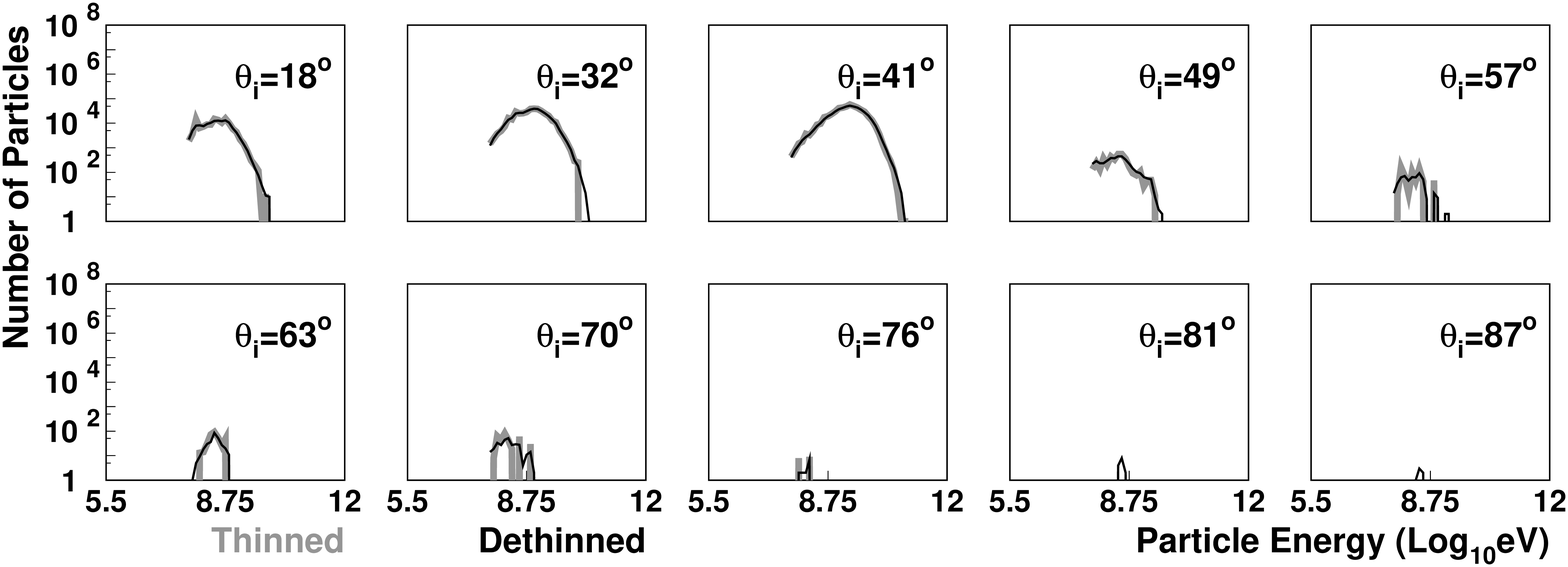}
\end{center}
\caption{Comparison of secondary particle spectra with and without dethinning 
for a thinned simulation of a protonic EAS with primary energy $E_0=10^{19.0}$~{eV} and primary zenith angle $\theta_0=45^\circ$.  In both cases, 
the secondary particles whose ground position was within
a region enclosed by shower rotation angles, $\Phi=[150^\circ,210^\circ]$ (with
respect to the primary azimuthal direction) and 
lateral distances, $r=[1500{\rm m},2000{\rm m}]$ were tabulated according
to particle type, incident angle with respect to the ground, $\theta_i$, and 
kinetic energy.  In the case of the thinned simulation, each secondary particle 
with weight, $w_i$, was treated as $w_i$ identical particles. 
The resulting spectral comparisons are shown in $\cos\theta_i=0.1$ increment
bins for a) photons, b) electrons, and c) muons. For each 
histogram, good agreement is observed between thinned simulations (gray) and 
dethinned (black).}

\label{fig:thdth2}
\end{figure*}

For the first comparison, we divide the particle flux into ten
$\cos\theta_i=0.1$ bins, where $\theta_i$ is the incident angle of an
individual particle
with respect to the ground.  Three particle types are considered: 
photons, electrons, and muons (all other types are relatively scarce).  Each bin 
is then histogrammed with respect to energy.  This results in $8\times 6\times 
10\times 3=1440$ discrete
secondary particle energy spectra.  By scanning through these spectra, 
any discrepancies in particle flux generated by dethinning can be readily 
identified.  Figures~\ref{fig:thdth1} and \ref{fig:thdth2} 
show examples, typical of all 1440 spectra, of these comparisons.  The agreement is excellent, showing that the process of smoothing the distribution of particles does not change the angular or energy distribution of particles from the parent thinned shower.

Having established that dethinning maintains the large-scale secondary 
particle fluxes from parent thinned simulations, we then turn to 
comparisons with non-thinned showers produced with a 
parallelized version of CORSIKA~\cite{parallel}.  Because it is structurally impossible in
CORSIKA to produce identical thinned and non-thinned simulations, for this comparison it is 
necessary to normalize the net particle flux of the non-thinned and thinned 
(not dethinned) simulations.  
This is done separately for each wedge-shaped region of the 
shower footprint and each particle type.    

\begin{figure*}[t,b,p]
\begin{center}
(a)\includegraphics[width=0.825\textwidth]{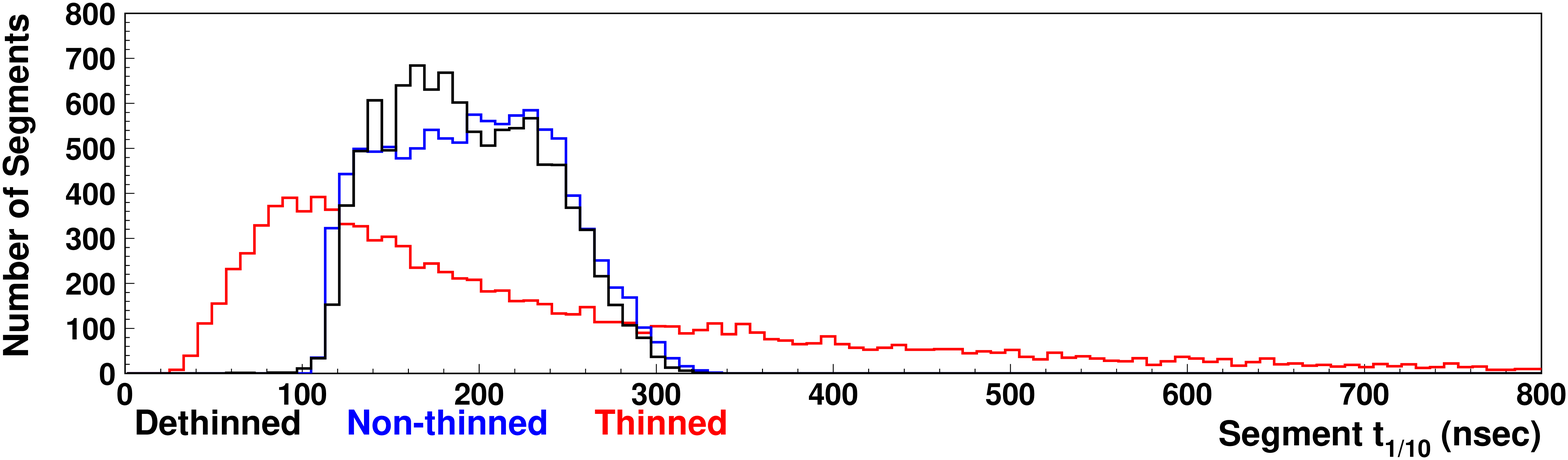}\\
(b)\includegraphics[width=0.825\textwidth]{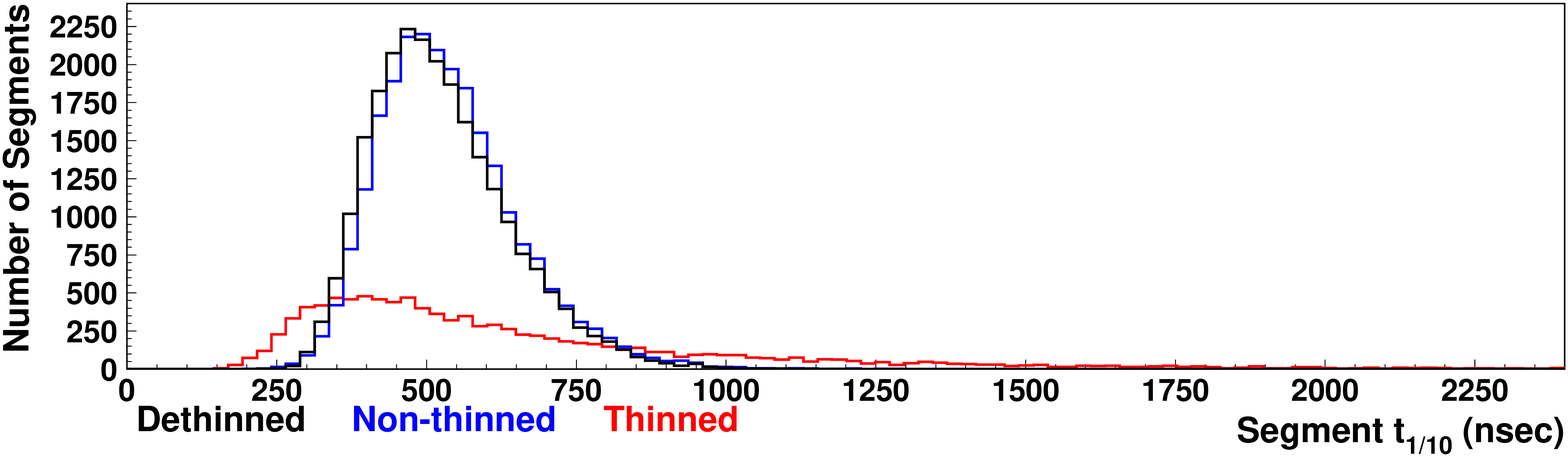}\\
\end{center}
\caption{Comparison of the distribution of rise times, $t_{1/10}$, where 10\% 
of the total particle flux has arrived for a given 
$6\times 6$~m$^2$ segment in plane normal to shower trajectory for $10^{19.0}$~eV 
protonic EAS simulations with a primary zenith angle of $45^\circ$.  For this
comparison, $t_{1/10}$ was measured for segments within
a) a region enclosed by shower rotation angles, $\Phi=[-30^\circ,30^\circ]$ 
(with respect to the primary azimuthal direction) and lateral distances, 
$r=[500{\rm m},1000{\rm m}]$  and b) a region enclosed by shower rotation 
angles, $\Phi=[150^\circ,210^\circ]$ (with respect to the primary azimuthal 
direction) and lateral distances, $r=[1500{\rm m},2000{\rm m}]$. In both cases, 
the distribution of $t_{1/10}$ values is consistent for the 
dethinned (black) and non-thinned (blue) simulations while the thinned 
(red) simulation is quite different.}
\label{fig:t0}
\end{figure*}

\begin{figure*}[t,b,p]
\begin{center}
(a)\includegraphics[width=0.825\textwidth]{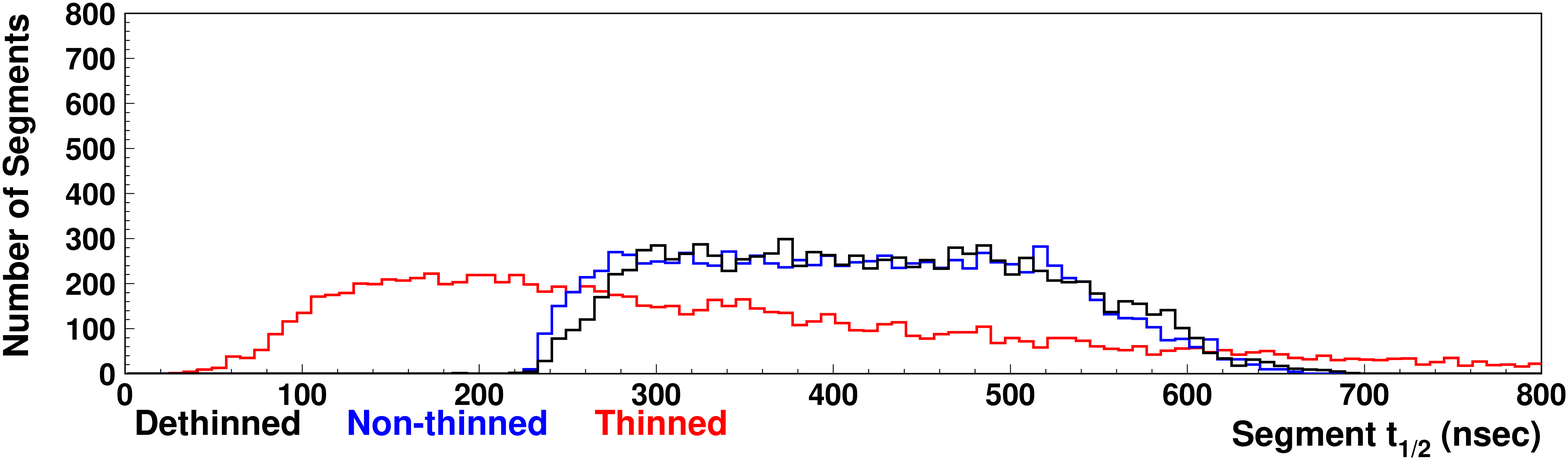}\\
(b)\includegraphics[width=0.825\textwidth]{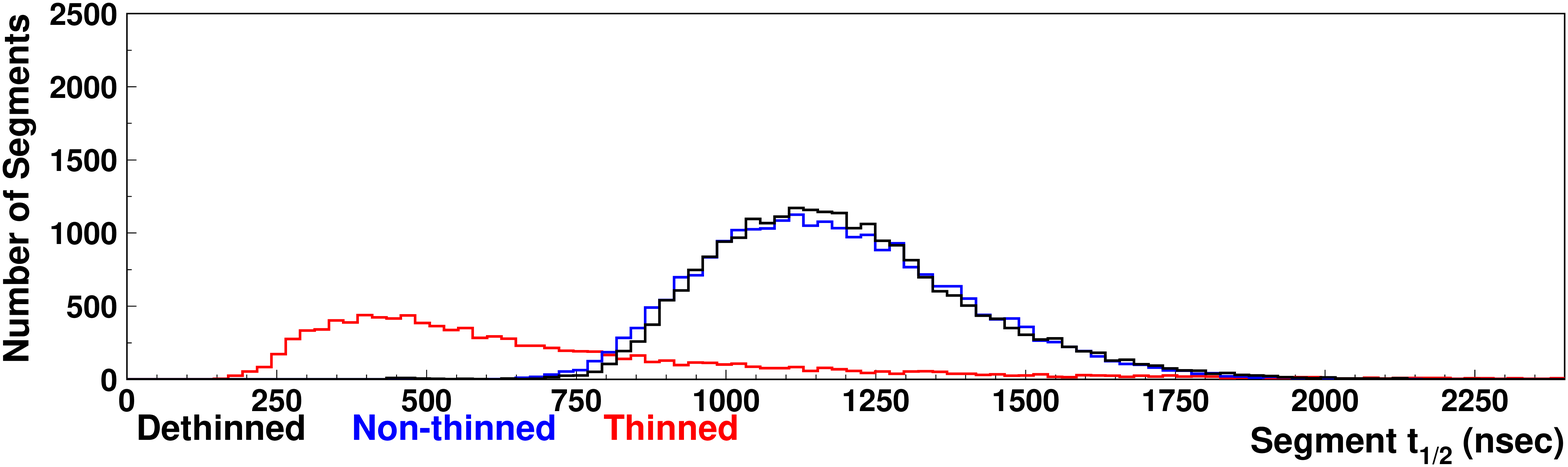}\\
\end{center}
\caption{Comparison of the distribution of median arrival times, 
$t_{1/2}$, where 
50\% of the total particle flux has arrived for a given  
$6\times 6$~m$^2$ segment in plane normal to shower trajectory for $10^{19.0}$~eV 
protonic EAS simulations with a primary zenith angle of $45^\circ$.  For this
comparison, $t_{1/2}$ was measured for segments within
a) a region enclosed by shower rotation angles, $\Phi=[-30^\circ,30^\circ]$ 
(with respect to the primary azimuthal direction) and lateral distances, 
$r=[500{\rm m},1000{\rm m}]$  and b) a region enclosed by shower rotation 
angles, $\Phi=[150^\circ,210^\circ]$ (with respect to the primary azimuthal 
direction) and lateral distances, $r=[1500{\rm m},2000{\rm m}]$. 
 In both cases, 
the distribution of $t_{1/2}$ values is consistent for the 
dethinned (black) and non-thinned (blue) simulations while the thinned 
(red) simulation is quite different.}
\label{fig:t05}
\end{figure*}

\begin{figure*}[t,b,p]
\begin{center}
(a)\includegraphics[width=0.825\textwidth]{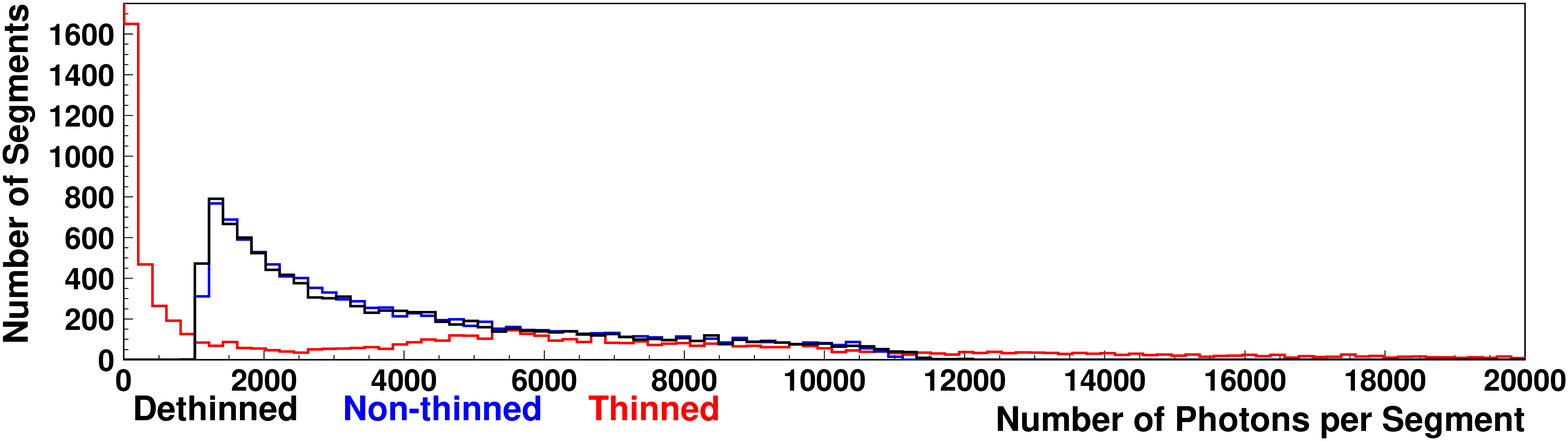}\\
(b)\includegraphics[width=0.825\textwidth]{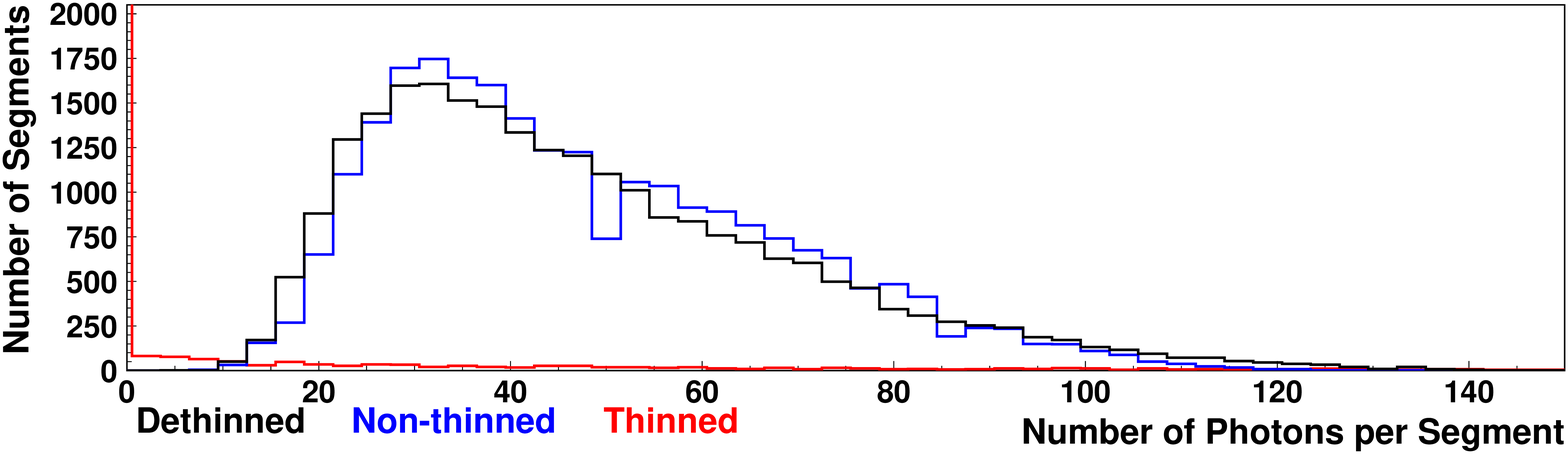}\\
\end{center}
\caption{Comparison of the distribution of photon flux measurements for 
$6\times 6$~m$^2$ segments in plane normal to shower trajectory for $10^{19.0}$~eV 
protonic EAS simulations with a primary zenith angle of $45^\circ$.  For this
comparison, photon flux measurements were done for segments within
a) a region enclosed by shower rotation angles, $\Phi=[-30^\circ,30^\circ]$ 
(with respect to the primary azimuthal direction) and lateral distances, 
$r=[500{\rm m},1000{\rm m}]$  and b) a region enclosed by shower rotation 
angles, $\Phi=[150^\circ,210^\circ]$ (with respect to the primary azimuthal 
direction) and lateral distances, $r=[1500{\rm m},2000{\rm m}]$. 
In both cases, 
the distribution of photon flux values is consistent for the 
dethinned (black) and non-thinned (blue) simulations while the thinned 
(red) simulation is quite different.}
\label{fig:gamma}
\end{figure*}

\begin{figure*}[t,b,p]
\begin{center}
(a)\includegraphics[width=0.825\textwidth]{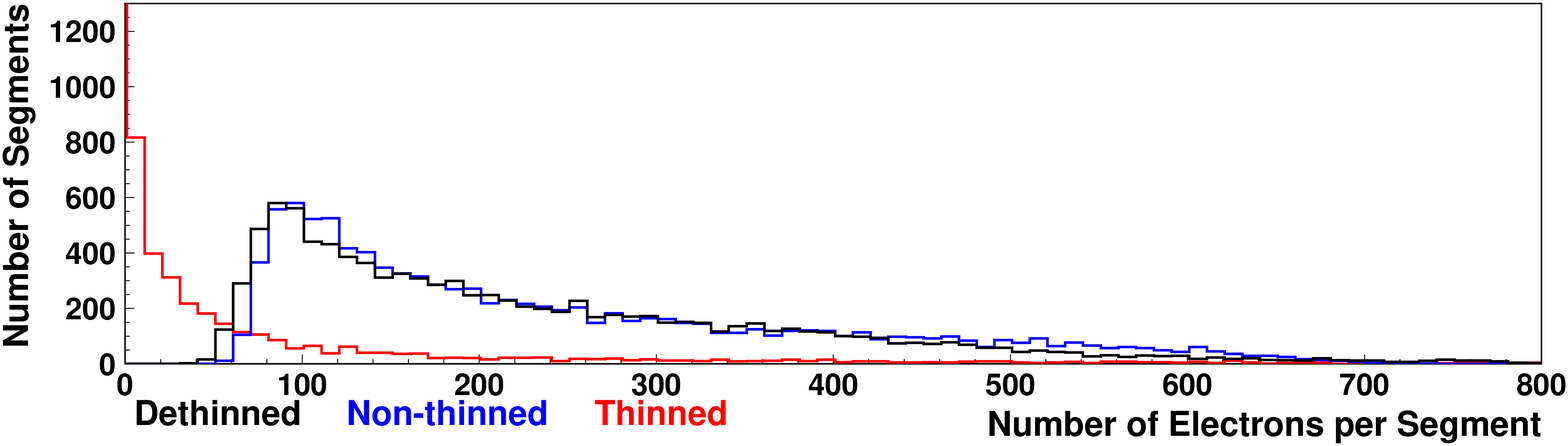}\\
(b)\includegraphics[width=0.825\textwidth]{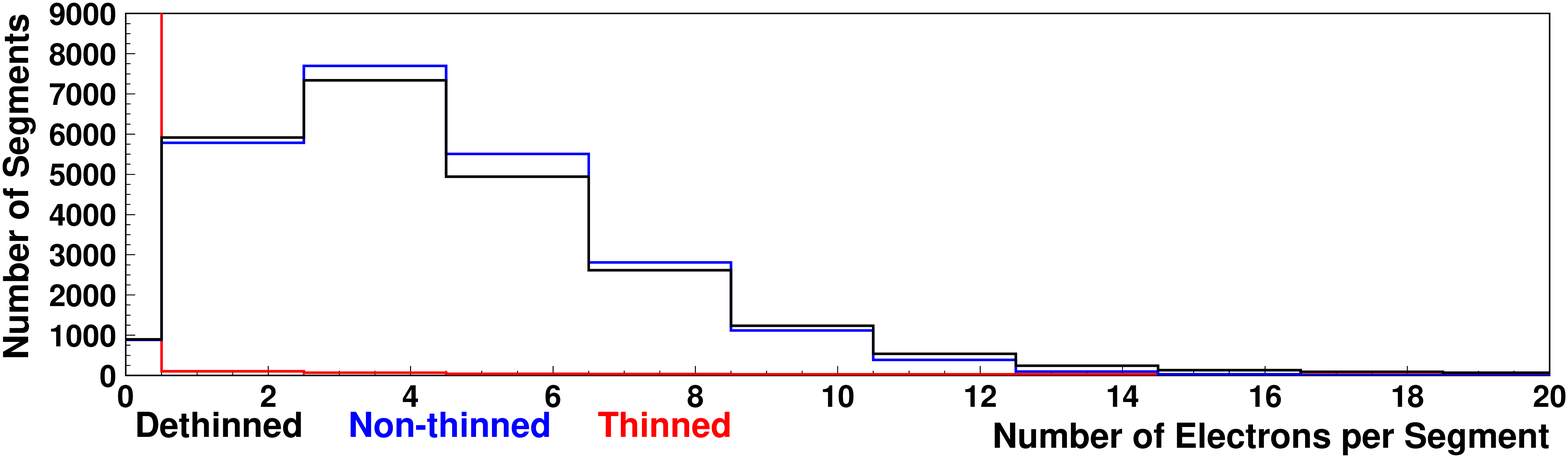}\\
\end{center}
\caption{Comparison of the distribution of electron flux measurements for 
$6\times 6$~m$^2$ segments in plane normal to shower trajectory for $10^{19.0}$~eV 
protonic EAS simulations with a primary zenith angle of $45^\circ$.  For this
comparison, electron flux measurements were done for segments within
a) a region enclosed by shower rotation angles, $\Phi=[-30^\circ,30^\circ]$ 
(with respect to the primary azimuthal direction) and lateral distances, 
$r=[500{\rm m},1000{\rm m}]$  and b) a region enclosed by shower rotation 
angles, $\Phi=[150^\circ,210^\circ]$ (with respect to the primary azimuthal 
direction) and lateral distances, $r=[1500{\rm m},2000{\rm m}]$. 
In both cases, 
the distribution of electron flux values is consistent for the 
dethinned (black) and non-thinned (blue) simulations while the thinned 
(red) simulation is quite different.}
\label{fig:electron}
\end{figure*}

\begin{figure*}[t,b,p]
\begin{center}
(a)\includegraphics[width=0.825\textwidth]{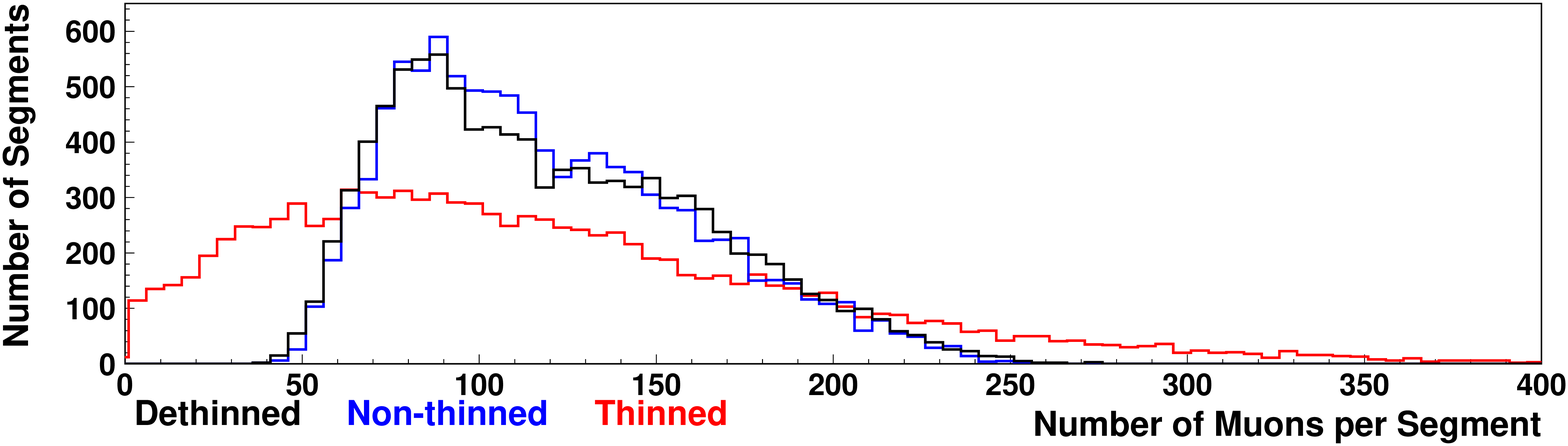}\\
(b)\includegraphics[width=0.825\textwidth]{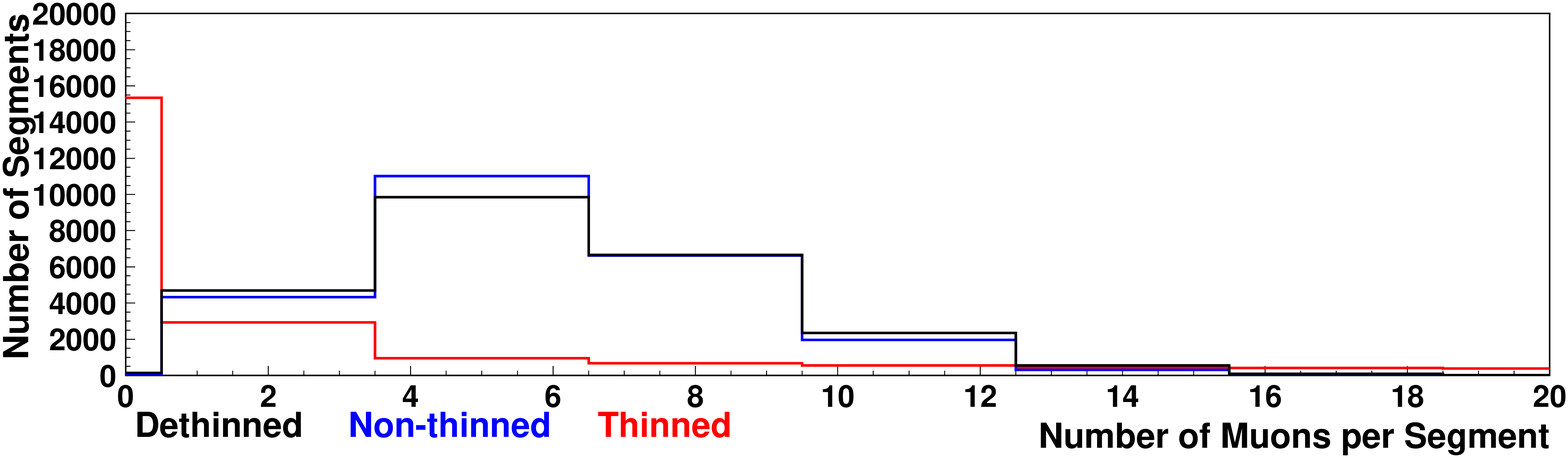}\\
\end{center}
\caption{Comparison of the distribution of muon flux measurements for 
$6\times 6$~m$^2$ segments in plane normal to shower trajectory for $10^{19.0}$~eV 
protonic EAS simulations with a primary zenith angle of $45^\circ$.  For this
comparison, muon flux measurements were done for segments within
a) a region enclosed by shower rotation angles, $\Phi=[-30^\circ,30^\circ]$ 
(with respect to the primary azimuthal direction) and lateral distances, 
$r=[500{\rm m},1000{\rm m}]$  and b) a region enclosed by shower rotation 
angles, $\Phi=[150^\circ,210^\circ]$ (with respect to the primary azimuthal 
direction) and lateral distances, $r=[1500{\rm m},2000{\rm m}]$. 
In both cases, 
the distribution of muon flux values is consistent for the 
dethinned (black) and non-thinned (blue) simulations while the thinned 
(red) simulation is quite different.}
\label{fig:mu}
\end{figure*}

For the comparison,  we consider the same segments in the shower footprint
as for Figures~\ref{fig:thdth1} and~\ref{fig:thdth2}. The segment is then further divided into $6\times 6$~m$^2$ tiles covering the distances from 500~m to 4500~m from the core.  
These tiles are then projected onto the ground, and for each 
tile we tabulate the time, $t_{1/10}$, 
when 10\% of the total particle flux has arrived, the time,
$t_{1/2}$, when 50\% 
of the total particle flux has arrived, and the flux of all
photons, electrons, and muons.  The times are
then corrected for the time offset between the positions of each segment on the 
ground and in the plane normal to the EAS. Figures~\ref{fig:t0} through
\ref{fig:mu}
show the results of this comparison.  They show that for the dethinned shower the time of arrival and number of particles per tile agree very well with that of the non-thinned shower.
 
\section{Conclusion}

The aim of this dethinning method is to use a thinned simulation to reconstruct, on a statistical basis, information lost in the thinning process.  Since thinning preserves mean particle densities as a function of radius from the shower core, but introduces large biases into the distribution of the densities (e.g., the RMS of the particle distribution), dethinning is designed to be a smoothing procedure.  The dethinning process is tuned to maintain mean particle densities from parent thinned showers, and reproduces distributions of arrival times and numbers of particles striking counter-size areas in non-thinned showers.  Since these are the distributions to which surface detectors of experiments are sensitive, dethinned showers can be used in place of non-thinned showers in comparisons with data. 

This method has three primary 
limitations. We require that $\varepsilon_{th} \leq 10^{-6}$ and lateral
distances be restricted to less than 4500~m from the shower core.  Beyond
these limits we cannot reliably control for artificial fluctuations.  We have tested this process for $\theta_0<60^\circ$ but have not yet examined the case of more inclined showers.

We have compared dethinned shower simulations against both thinned and non-thinned simulations, and shown that the dethinning process reproduces the characteristics of CORSIKA/QGSJET-II-03 showers generated without thinning.  In a future paper in this series we will show that the resulting showers reproduce the characteristics of the TA surface array data.  

Dethinning is proving to be a powerful tool for studying and understanding UHECR 
observations by enabling a thorough simulation of surface array data.  This enables a direct comparison between Monte Carlo simulations and TA surface array data and results in a more complete understanding of the response of the detector.  This understanding leads to the ability to accurately assess detector aperture for efficiencies far less than 100\%, which in turn leads to significant improvements in the measurement of the cosmic ray spectrum, and in the estimation of the detector exposure to the sky for astronomical studies.

\section{Acknowledgments}

This work was supported by the U.S.~National Science Foundation awards 
PHY-0601915, PHY-0703893, PHY-0758342, and PHY-0848320 (Utah) 
and PHY-0649681 (Rutgers). 
The simulations presented herein have only been possible due to 
the availability of computational resources at the Center for High Performance 
Computing at the University of Utah which is gratefully acknowledged. A portion
of the 
computational resources for this project has been provided by the U.S.~National 
Institutes of Health (Grant \# NCRR 1 S10 RR17214-01) on the Arches 
Metacluster, administered by the University of Utah Center for High 
Performance Computing.

\bibliographystyle{elsarticle-num}
\bibliography{dethinning}

\end{document}